\documentclass[aps,preprint,nofootinbib,showpacs]{revtex4}

\usepackage{graphicx}
\usepackage{amsmath}
\usepackage{amssymb}
\usepackage{epsf,latexsym}
\usepackage{times}
\usepackage{epsfig}
\usepackage{citesort}

\def\be{\begin{equation}}
\def\ee{\end{equation}}
\def\bea{\begin{eqnarray}}
\def\eea{\end{eqnarray}}

\def\lsim{\raise0.3ex\hbox{$\;<$\kern-0.75em\raise-1.1ex\hbox{$\sim\;$}}}
\def\gsim{\raise0.3ex\hbox{$\;>$\kern-0.75em\raise-1.1ex\hbox{$\sim\;$}}}

\begin{document}

\title{Sterile neutrino dark matter in $B-L$ extension of the standard model and galactic $511$ keV line}

\author{Shaaban Khalil}
\email{skhalil@bue.edu.eg}
\affiliation{Centre for Theoretical
Physics, The British University in Egypt, El Sherouk City, Postal
No. 11837, P.O. Box 43, Egypt.} %
\affiliation{Department of Mathematics, Ain Shams University,
Faculty of Science, Cairo, 11566, Egypt.}

\author{Osamu Seto \footnote{Present address: William I. Fine Theoretical Physics Institute, University of Minnesota, Minneapolis, MN 55455, USA}}
\email{seto@physics.umn.edu}
\affiliation{
 Instituto de F\'{i}sica Te\'{o}rica UAM/CSIC,
 Universidad Aut\'{o}noma de Madrid, Cantoblanco, Madrid 28049, Spain.}

\date{\today }

\begin{abstract}
Sterile right-handed neutrinos can be naturally embedded in a low
scale gauged $U(1)_{B-L}$ extension of the standard model. 
We show that, within a low reheating scenario, such a neutrino is an interesting
dark matter candidate. We emphasize that if the neutrino mass is of the
order of MeV, then it accounts for the measured dark matter relic density and 
also accommodates the observed flux of 511 keV photons from the galactic bulge.
\end{abstract}


\preprint{IFT-UAM/CSIC-08-19}

\maketitle

%
\section{Introduction}

The existence of non-baryonic dark matter (DM) and non-vanishing
 neutrino masses~\cite{NeutrinoReview}, based on the observation
 of neutrino oscillation, are considered as the most important
evidences for new physics beyond the standard model (SM). Here, we
argue that these two exciting problems may be simultaneously
solved in a simple extension of the SM, that is based on the gauge
group: $G_{B-L} = SU(3)_C \times SU(2)_L \times U(1)_Y \times
U(1)_{B-L}$~\cite{B-L,khalil}. In this class of models, three SM
singlet fermions arise naturally
 due to the $U(1)_{B-L}$ anomaly cancellation condition.
These singlet fermions are regarded as the right-handed
 neutrinos~\cite{khalil}. The scale of the
right-handed neutrino masses is no longer arbitrary, it is
proportional to the scale of $B-L$ symmetry breaking. 
It was shown that, similarly to the electroweak symmetry, the scale of $B-L$
can be linked with the supersymmetry breaking scale at the
observed sector and then the $B-L$ symmetry is radiatively
broken at TeV scale~\cite{masiero}.

This model of low scale $B-L$ extension of the SM can account for
the experimental results for the small neutrino masses and their
large mixing through a low scale seesaw mechanism~\cite{Seesaw}
and Dirac neutrino Yukawa couplings of order $\lsim
10^{-6}$~\cite{Abbas:2007ag}. 
Moreover, this model predicts an extra neutral gauge boson $Z'$
 corresponding to the $B-L$ gauge symmetry and 
 an extra SM singlet scalar (extra Higgs). 
The phenomenology of this model and its potential discovery at 
 the Large Hadron Collider (LHC) has been 
 studied~\cite{Emam:2007dy,Huitu:2008gf}.

As emphasized in~\cite{khalil}, a right-handed
neutrino $\nu_R$ has tree level interactions with the SM leptons $L$, SM
Higgs $\phi$, $Z'$, and extra Higgs $\chi$. They are described by
\begin{equation}
{\cal L}^{{\rm int}}_{\nu_R} = q_{\nu_R} g_{B-L} \bar{\nu}_R
Z'_{\mu} \sigma^{\mu} \nu_R - ( \lambda_{\nu} \bar{L} \tilde{\phi}
{\nu}_R
 +\frac{1}{2}\lambda_{\nu_R} \bar{\nu}_R^c \chi \nu_R +
 {\rm h.c.}).
\label{lagrangian}
\end{equation}
The first term is due to the covariant derivative for right-handed
neutrinos, $g_{B-L}$ is the $B-L$ gauge coupling and $q_{\nu_R}$ is its charge.
Other interaction terms refer to the possible Yukawa interactions that
involve the right-handed neutrinos. After the symmetry breakdown,
the right-handed neutrinos acquire Majorana masses
\begin{eqnarray}
M_{\nu_{R_i}}= \lambda_{\nu_{R_i}} \langle \chi\rangle
 \equiv \frac{1}{\sqrt{2}} \lambda_{\nu_{R_i}} v^{\prime}.
\end{eqnarray}
For $v^{\prime}$ of $ {\cal O} (1)$ TeV,
 the right-handed neutrino masses may vary, according to the values of the Yukawa couplings $\lambda_{\nu_{R_i}}$, from TeV scale to a lighter scale.
If the right-handed neutrinos are subjected to the flavour symmetry that
 should account for the quark and lepton mass hierarchy, then it is
 very natural to find $M_{\nu_{R_1}} \ll M_{\nu_{R_2}} <
 M_{\nu_{R_3}}$.
In this case, it becomes quite plausible that the mass of
 the lightest right-handed neutrino is of order of 
 MeV~\footnote{The smallness might be a consequence of an extra dimension
 ~\cite{Kadota}}.

In this paper, we consider the scenario where a right-handed neutrino
 can be a DM candidate with the sufficient abundance.
In addition, it provides a natural explanation for the flux of 511 keV photons 
 from the galactic bulge observed by INTEGRAL satellite~\cite{spi}.
It was suggested that this emission might be originated from
 the annihilation~\cite{Boehm:2003bt,Fayet:2004bw,Gunion:2005rw,Boehm:2006gu,Hooper:2008im}
 or decay~\cite{Hooper:2004qf,Picciotto:2004rp,Kawasaki:2005xj,Takahashi:2005kp,Kasuya:2005ay,Kasuya:2006kj,Chun:2006ss,Pospelov:2007xh}
 of DM particles into low energy electron-positron pairs.
The outgoing positron loses its kinetic energy by collisions with baryonic
 materials and eventually forms positronium with an electron in the medium.
This positronium would annihilate into monoenergetic photons of 511 keV. 
Therefore, it was concluded that the DM candidates have to be quite light, 
 perhaps as light as electron; 
 otherwise the positrons produced are too energetic to form positronium.
Conservative constraints imply that the mass of the corresponding DM
 should be of ${\cal O}(10)$ MeV.

The paper is organized as follows. 
In section 2 we show that the lightest right-handed neutrinos can be produced
 in enough quantities to be the main component of dark matter 
 in a thermal history with order MeV reheating temperature. 
In section 3 we discuss the possibility that
 the MeV right-handed neutrino in $B-L$ extension of the SM can be a source
 for the observed $511$ keV gamma ray line. 
Finally we give our conclusion and summary remarks in section 4.

\section{dark matter}

In this section we consider the lightest right-handed neutrino in
TeV scale $B-L$ extension of the SM as a DM candidate. 

First, one should notice that this dark matter candidate is not stable.
This is because, to be precise, the dark matter candidate is the mixture of 
 a left-handed neutrino $\nu_L$ and a right-handed neutrino $\nu_R$ 
 in the mass eigen states obtained by diagonalizing the mass matrix
\be 
M(\nu_L,\nu_R) = \left(\begin{array}{cc}
 0 & m_D^T \\
 m_D & M_{\nu_R}
 \label{Mnu}
\end{array}\right),
\ee
 where $m_D$ is the Dirac neutrino mass term. 
Although the lightest right-handed-like neutrino $N_1$ 
 is a candidate of dark matter,
 it contains a tiny component of a left-handed neutrino that interacts
 through the weak gauge bosons $W^{\pm}$ and $Z$.
Thus, even if $N_1$ is ligher than the SM-like
 Higgs, it may still decay into lepton, anti-lepton and neutrino
 via $Z$ and $W$ exchange. 
We will examine this decay in detail in the Sec.~4.


Hence, the sterile (right-handed) neutrino can be considered as DM
candidate if it is rather light, around MeV or less, and if the
mixing with light neutrinos is very tiny, in order to have a
sufficiently long lifetime. 
Furthermore, the radiative decay $N_1 \rightarrow \nu + \gamma$ often gives
 further stringent constraints, for keV $-$ MeV sterile neutrinos, on the 
 left-right mixing due to the line $\gamma$-ray and x-ray background 
 radiation~\cite{Boyarsky1,Boyarsky2}. 
Therefore, from now on, we assume that the DM candidate right-handed neutrino
 has a small enough mixing with left-handed neutrino 
 to satisfy these constraints.

It has been assumed that
 sterile neutrino dark matter production is realized by
 so-called Dodelson-Widrow (DW) mechanism~\cite{Dodelson:1993je}
 due to the mixing oscillation
 between active and sterile neutrinos~\cite{Dodelson:1993je,Dolgov:2000ew}.
However, the recent stringent bound from x-ray background seems to exclude 
 such unstable sterile neutrino dark matter with 
 the minimal production scenario via DW mechanism~\cite{Asaka,Yuksel:2007xh}.
To be precise, for a small enough active-sterile mixing to be consistent
 with the x-ray background constraints,
 the abundance of sterile neutrino produced by DW mechanism is too low
 to account for whole dark matter~\cite{Palazzo:2007gz}.
Hence, another production mechanism is necessary
 if one wants to construct a sterile neutrino
 as the principal component of dark matter.
For instance, Shaposhnikov and Tkachev~\cite{Shaposhnikov}, and 
 Kusenko~\cite{Kusenko,Petraki:2007gq}
 proposed the production of sterile neutrinos by the decay of
 new gauge singlet scalar particle.
This is also one fact leading us to introduce a new interaction
 to produce sterile neutrinos.

It is remarkable that $N_1$ in this model can not be thermal
relics. $N_1$ has the gauge interaction of $U(1)_{B-L}$ with
 apparently heavy $Z'$ exchange
 \footnote{$Z'$ can be very light, if $g_{B-L}$ is also very small.}.
Then, the thermal relic abundance is qualitatively same as
 that of heavy left-handed neutrinos.
Here, $Z'$ essentially plays the role of the weak boson
 in the standard~\cite{LeeWeinberg,KolbTurner}. 
If such a light $N_1$ is thermal relic, it becomes hot dark matter. 
Therefore, we will consider a sort of non-thermal production
 under a non-standard thermal history with a low reheating temperature $T_R$.
Such a thermal history may be realized by some dominated late decaying
 object such as a moduli field or
 a late mini inflation~\cite{Giudice:2000ex,Khalil:2002mu}.
The studies of sterile neutrino dark matter under
 such a thermal history were also done~\cite{Gelmini,Yaguna:2007wi,Gelmini:2008fq}.

The leading diagram for $N$ production is illustrated in figure 1. 
As can be seen from the diagram, in $B-L$ extension of the SM, 
 the production of MeV right-handed neutrino is dominated by
 the $Z'$ exchange contribution from leptons and left-handed neutrinos.
The effects of intermediate $H'$ and $H$ scalar Higgs bosons are strongly
 suppressed and can be safely neglected.
\begin{figure}[h,t]
\begin{center}
\epsfig{file=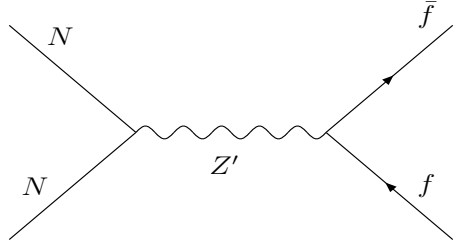, width=6cm,height=3.25cm,angle=0}
\end{center}
\caption{Feynman diagrams in $B-L$ extension of the SM that dominantly 
contribute to the right-handed neutrino production by the $Z'$ exchange.}
\label{fig1}
\end{figure}

In the production, a small $\nu_L$ component in $N_1$ is not important at all,
 since the oscillation production is not efficient in our scenario.
Hence, in the rest of this section, we consider $N_1 = N_R$, where 
 $N_R$ denotes a right-handed neutrino in 
 the four component Majorana representation $N_R= \nu_R + \nu_R^c$
 which satisfies $N_R^c = N_R$.

The relic abundance can be estimated by solving the Boltzmann equation
\begin{equation}
 \dot{n}+3 H n =  \langle\sigma v(ij \rightarrow NN)\rangle n_i n_j .
\end{equation}
$n$ is the number density of $N$. $\langle\sigma v\rangle$
 is the thermally averaged product of the pair production cross section of $N$
 and the relative velocity with the initial state particles $i$ and $j$.
For the calculation of the production cross section of right-handed neutrinos
 from the initial state $i$ and $j$ particles, we introduce $w(s)$ defined by
\begin{eqnarray}
w(s; ij \rightarrow) \equiv \frac{1}{32 \pi} \int \frac{d \cos\theta}{2}
 \sqrt{\frac{s-4 m_{\rm in}^2}{s}} |{\bar {\cal M}_{ij}}(s, \cos\theta)|^2,
\end{eqnarray}
 where $s$ is the usual Mandelstam variable,
 $\theta$ is the scattering angle in the center of mass frame,
 $m_{\rm in}$ is the mass of the initial state particle,
 and $|{\bar {\cal M}}|^2$ is the squared amplitude matrix
 after the spin summing and averaging and also some multiplications
 for identical particles.
Then, the right-hand side of the Boltzmann equation can be rewritten as
\begin{eqnarray}
  \langle\sigma v(ij\rightarrow NN)\rangle n_i n_j
  =  \frac{4 T^4}{(2\pi)^5}\int_{\frac{2m_N}{T}}^{\infty}
 dx \, x^2 K_1(x) \sum_{(i,j)} g_i g_j w(x^2 T^2; ij\rightarrow) ,
\end{eqnarray}
 with $x \equiv \sqrt{s}/T$ and
 $K_1(x)$ being the first order modified Bessel function.
Here $g_i$ stand for the internal degrees of freedoms of the $i$ particle.

As we will show, the suitable reheating temperature is about several MeV.
Hence, the relevant initial states for the production are
 $\nu \, {\bar \nu}$ and $e^+ e^-$. For each initial state, we found
\begin{eqnarray}
&& \int \frac{d \cos\theta}{2}|{\bar{\cal M}_{\nu {\bar \nu}}}(s, \cos\theta)|^2
 = \frac{2}{3} \frac{|g_{B-L}^2 q_N q_{\nu}|^2 }{(s-M_{Z'}^2)^2 + M_{Z'}^2
\Gamma_{Z'}^2} (s-4m_N^2)s ,\\
&& \int \frac{d \cos\theta}{2}|{\bar{\cal M}_{e^+ e^-}}(s, \cos\theta)|^2
  = \frac{4}{3} \frac{ |g_{B-L}^2 q_N q_e|^2 }{ (s-M_{Z'}^2)^2 +
M_{Z'}^2\Gamma_{Z'}^2} (s-4m_N^2) (s+2 m_e^2) ,
\end{eqnarray}
 where $q_N=q_e=q_{\nu}=-1$ are the $B-L$ charges respectively.
The gauge boson $Z'$ has the mass $M_{Z'}$,
 would be of order ${\cal O}(100)$ GeV for $g_{B-L} = {\cal O}(0.1)$,
 and its decay width $\Gamma_{Z'}$.

By introducing $Y \equiv n/s$ with $s$ being the entropy density
 and using the relation $\dot{T}=-HT$, we found the final abundance as
\begin{eqnarray}
Y_{\infty} &=& \int^{T_R} dT
 \frac{\langle\sigma v(ij\rightarrow NN)\rangle n_i n_j }{s H T} dT \nonumber\\
&\simeq& 3.5 \times 10^{-7} \left(\frac{10}{g_*}\right)^{3/2}
\left(\frac{T_R}{5 \,{\rm MeV}}\right)^3 \left(\frac{9.7 \,
 {\rm TeV}}{M_{Z'}/g_{B-L}}\right)^4 ,
\label{Y:N}
\end{eqnarray}
 for our range of interest $M_{Z'} \gg T_R \gg m_N $.
This can be rewritten in terms of the density parameter as 
\be
\Omega_N h^2 \simeq 0.1 \times
\left(\frac{m_N}{1 \,{\rm MeV}}\right)
\left(\frac{Y_{\infty}}{3.5 \times 10^{-7}}\right).
\label{Omegah^2:N}
\ee
One can see from this equation that the heavier $N$, the larger
$\Omega_N h^2$ obtained. For given $T_R$ and $M_{Z'}/g_{B-L}$,
if the mass of $N_i$ is larger than
 the value corresponding to $\Omega_N h^2=0.1$ and lower than $T_R$
\footnote{Needless to say, for $m_{N_i} > T_R$,
 such a heavy neutrino production is kinematically blocked.},
 then such heavier $N_i$ has to be unstable and decay
 in order not to behave as dark matter.

For $m_N \simeq 1$ MeV,
 the experimental lower bound $M_{Z'}/g_{B-L} \gtrsim 6$ TeV corresponds to
 the lower bound on the reheating temperature $T_R \gtrsim 2.7$ MeV,
 form equation~(\ref{Y:N}).
Since the lower bound is about $2$ MeV~\cite{LowerBoundOnTR},
 the consistency with the big bang nucleosynthesis is ensured
 for $m_N \sim 1$ MeV.

Finally, one may notice that the neutrinos produced have a thermal spectrum
 though they are not thermal relic, in other words, they were not in equilibrium,
 because the initial state particles in the scattering processes of
 the production, neutrinos and $e^+, e^-$, are in thermal equilibrium.
This is simply same as the thermal production of gravitino dark matter
 by scatterings in a thermal bath~\cite{Bolz}.
Hence, if the right-handed neutrinos are light enough,
 they could be warm dark matter and the mass would be constrained
 as $m_N \gtrsim 10$ keV by the Lyman alpha, as usual.

\section{INTEGRAL anomaly}

We now consider the possibility that our right-handed neutrino
 with $\cal{O}$(MeV) mass in $B-L$ extension of the SM can account for
 the observed flux of 511 keV photons.
As mentioned above, the SPI spectrometer on the INTEGRAL satellite has detected
 an intense $511$ keV gamma ray line flux which, at $2 \sigma$ level,
 is given by~\cite{spi}:
\be %
\Phi_{\gamma, 511} = (1.05 \pm 0.06) \times 10^{-3} {\rm photon}
~{\rm cm}^{-2} s^{-1} , %
\ee
with a width of 3 keV.

The flux of the gamma ray may be explained through the decay of
 right-handed-like neutrino dark matter.
The interaction is given by
\begin{eqnarray}
{\cal L}_{\rm int}
 = \bar{\nu}_i (U_{\alpha i})^{\dagger} U_{\alpha I}\gamma^{\mu}
  \left(\frac{g}{2 \cos\theta_W}Z_{\mu} + g_{B-L}q_{\nu} Z'_{\mu}\right)N_I
  + \bar{l^-}_{\alpha}\gamma^{\mu}\frac{g}{\sqrt{2}}W^- U_{\alpha I} N_I
  + {\rm  h.c.},
\end{eqnarray}
 where $\nu_i$ and $N_I$ are left- and right-handed-like neutrino
 in the mass eigen state respectively. 
Both $i$ and $I$ run over $1, 2, 3$.
$\theta_W$ is the Weinberg angle, and
 $U$ is the unitary matrix for transforming neutrinos from 
 the flavour and Majorana mass eigenstate into the actual mass eigenstate 
 with $\alpha$ being left-handed flavour and
 right-handed Majorana mass index, in other words,
 $\alpha$ runs $e, \mu, \tau, R_1, R_2, R_3$.
The rate of decay of dark matter $N_1$ into $e^+, e^-, \nu$ is
estimated as
\begin{eqnarray}
\Gamma(N_1 \rightarrow e, \bar{e}, \nu)
 &\simeq& \frac{G_F^2 m_N^5}{192 \pi^3}
 \sum_i \left| \sum_{\alpha} (U_{\alpha i})^{\dagger} U_{\alpha 1}\right|^2
 \nonumber \\
 && \times \left[\frac{c_A^2}{2}
 + \frac{(- c_A)}{2}\left(\frac{(U_{e i})^{\dagger} U_{e N_1}}{\sum_{\alpha} (U_{\alpha i})^{\dagger} U_{\alpha N_1}} + {\rm c.c.}\right)
 + \left|\frac{(U_{e i})^{\dagger} U_{e N_1}}
 {\sum_{\alpha} (U_{\alpha i})^{\dagger} U_{\alpha N_1}}\right|^2 \right] ,
\label{DecayRate}
\end{eqnarray}
 in the $m_e \ll m_N$ limit, with $G_F$ being the Fermi constant.
Here, the $Z'$ gauge boson exchange contribution is omitted
 because it is much smaller than the others.
In addition, the vector contribution of the neutral current is neglected because
 $(c_V, c_A) = ({\cal O}(0.01), -1/2)$ for the electron.
We can regard the factor
 $\sum_i \left| \sum_{\alpha} (U_{\alpha i})^{\dagger} U_{\alpha 1}\right|^2 $
 as $\sin^2 2\theta/4$ in the standard notation
 for the mixing of a sterile neutrino.

It has been shown that decaying dark matter produces the 511 keV gamma ray
 flux of~\cite{Hooper:2004qf,Picciotto:2004rp}
\begin{equation}
\Phi_{511 \gamma} \sim 10^{-3} \left( \frac{10^{27} \,{\rm s}}
 {\Gamma(N_1 \rightarrow e\bar{e}\nu)^{-1}}\right)
 \left( \frac{1 \,{\rm MeV}}{m_N}\right)  {\rm cm}^{-2} s^{-1}.
\label{Flux}
\end{equation}
From equations~(\ref{DecayRate}) and (\ref{Flux}) with
 $((U_{e i})^{\dagger} U_{e N_1})
 /(\sum_{\alpha} (U_{\alpha i})^{\dagger} U_{\alpha N_1})  \leq 1$,
 one finds that the observed 511 keV line gamma ray can be explained if the mixing
 angle is of order:
\begin{equation}
 \sin^2 2\theta \left(\frac{m_N}{1 {\rm MeV}}\right)^4
  \simeq  10^{-22} ,
\end{equation}
which is consistent with the small mixing between light and heavy
neutrinos in such models.

\section{Summary}

We pointed out a new mechanism of production of sterile neutrino
 by new $U(1)_{B-L}$ gauge boson $Z'$
 under a kind of non-standard thermal history with a low reheating temperature.
If the reheating temperature is low enough,
 even if there is an additional gauge interaction,
 sterile neutrinos cannot reach thermal equilibrium.
Hence, these are produced only through the scattering in the thermal bath.
By adjusting the reheating temperature $T_R$ and
 the scale of $B-L$, $M_{Z'}/g_{B-L}$,
 sufficient amounts of sterile neutrinos can be produced 
 for rather wide range of mass.

In addition, we also have shown it can account for 511 keV gamma ray
 from galactic bulge claimed by INTEGRAL/SPI, for $m_N \gtrsim 1$ MeV.
Picciotto and Pospelov have also explained this via decaying sterile neutrino.
However, in their model, sterile neutrino cannot be abundant enough
 to account for whole dark matter energy density
 since they assumed a DW mechanism for the production~\cite{Picciotto:2004rp}.
This is because the left-right mixing angle is responsible for both
 the production and the decay of dark matter $\nu_R$.
For given $\sin^2 2 \theta$ suitable for sufficient production
 of dark matter via the DW mechanism, the predicted flux turned out too strong.
Thus, another additional dark matter component is needed to make up for
 the lack of the dark matter energy density.
Some other models also require additional species of dark matter,
 in other words multi-component dark matter,
 to obtain $\Omega_{DM} h^2 \simeq 0.1$ ~\cite{Kawasaki:2005xj,Kasuya:2005ay}
 for a similar reason, namely too strong flux.
In contrast, in our scenario,
 a single component of dark matter is enough to explain
 both the cosmological abundance and the flux of 511 keV line gamma ray.
On the other hand,
 the price to pay is to assume a low reheating temperature scenario, 
 where successful baryogenesis is hard 
 in general, though Affleck-Dine mechanism~\cite{Affleck:1984fy} 
 might be applicable~\cite{Dolgov:2002vf} 
 if we supersymmetrize the model as in~\cite{masiero}.

%
\section*{Acknowledgments}
OS is grateful for warm hospitality at Centre for Theoretical Physics 
 in The British University in Egypt where this work was initiated.
The work of OS is supported by the MEC project FPA 2004-02015,
 the Comunidad de Madrid project HEPHACOS (No.~P-ESP-00346).
The work of SK was partially supported by the ICTP grant project 30
 and the Egyptian Academy of scientific research and technology.



\begin{thebibliography}{99}

\bibitem{NeutrinoReview}
For a review, see e.g.,
  R.~N.~Mohapatra {\it et al.}, Rept.\ Prog.\ Phys.\  {\bf 70}, 1757 (2007).

\bibitem{B-L}
  R.~N.~Mohapatra and R.~E.~Marshak, Phys.\ Rev.\ Lett.\  {\bf 44}, 1316 (1980) [Erratum-ibid.\  {\bf 44}, 1643 (1980)];
  R.~E.~Marshak and R.~N.~Mohapatra, Phys.\ Lett.\  B {\bf 91}, 222 (1980);
  C.~Wetterich, Nucl.\ Phys.\  B {\bf 187}, 343 (1981).

\bibitem{khalil}
  S.~Khalil,
  J.\ Phys.\ G. {\bf 35} 055001 (2008).

\bibitem{masiero}
  S.~Khalil and A.~Masiero,
  Phys.\ Lett.\  B {\bf 665}, 374 (2008).   

\bibitem{Seesaw}
T.~Yanagida,
 in \textit{Proceedings of Workshop on the Unified Theory and
 the Baryon Number in the Universe}, Tsukuba, Japan,
 edited by A.~Sawada and A.~Sugamoto (KEK, Tsukuba, 1979), p 95;
M.~Gell-Mann, P.~Ramond, and R.~Slansky,
 in \textit{Supergravity},
 Proceedings of Workshop, Stony Brook, New York, 1979, edited by
 P.~Van~Nieuwenhuizen and D.~Z.~Freedman
 (North-Holland, Amsterdam, 1979), p 315;
R.~N.~Mohapatra and G.~Senjanovic, Phys. Rev. Lett. {\bf 44}, 912 (1980).

\bibitem{Abbas:2007ag}
  M.~Abbas and S.~Khalil,
  JHEP {\bf 0804}, 056 (2008).

\bibitem{Emam:2007dy}
  W.~Emam and S.~Khalil,
  Eur.\ Phys.\ J.\  C {\bf 522}, 625 (2007).

\bibitem{Huitu:2008gf}
  K.~Huitu, S.~Khalil, H.~Okada and S.~K.~Rai,
  arXiv:0803.2799 [hep-ph].
  
\bibitem{Kadota}
K.~Kadota, Phys.\ Rev.\  D {\bf 77}, 063509 (2008).

\bibitem{spi}
J.~Knodlseder {\it et al.},
  Astron.\ Astrophys.\  {\bf 411}, L457 (2003);  
P.~Jean {\it et al.},
  Astron.\ Astrophys.\  {\bf 407}, L55 (2003).

\bibitem{Boehm:2003bt}
  C.~Boehm, D.~Hooper, J.~Silk, M.~Casse and J.~Paul,
  Phys.\ Rev.\ Lett.\  {\bf 92}, 101301 (2004).

\bibitem{Fayet:2004bw}
  P.~Fayet,
  Phys.\ Rev.\  D {\bf 70}, 023514 (2004).

\bibitem{Gunion:2005rw}
  J.~F.~Gunion, D.~Hooper and B.~McElrath,
  Phys.\ Rev.\  D {\bf 73}, 015011 (2006).

\bibitem{Boehm:2006gu}
  C.~Boehm, J.~Orloff and P.~Salati,
  Phys.\ Lett.\  B {\bf 641}, 247 (2006).

\bibitem{Hooper:2008im}
  D.~Hooper and K.~M.~Zurek,
  Phys.\ Rev.\  D {\bf 77}, 087302 (2008).   

\bibitem{Hooper:2004qf}
  D.~Hooper and L.~T.~Wang,
  Phys.\ Rev.\  D {\bf 70}, 063506 (2004).

\bibitem{Picciotto:2004rp}
  C.~Picciotto and M.~Pospelov,
  Phys.\ Lett.\  B {\bf 605}, 15 (2005).

\bibitem{Kawasaki:2005xj}
  M.~Kawasaki and T.~Yanagida,
  Phys.\ Lett.\  B {\bf 624}, 162 (2005).

\bibitem{Takahashi:2005kp}
  F.~Takahashi and T.~T.~Yanagida,
  Phys.\ Lett.\  B {\bf 635}, 57 (2006).

\bibitem{Kasuya:2005ay}
  S.~Kasuya and F.~Takahashi,
  Phys.\ Rev.\  D {\bf 72}, 085015 (2005).

\bibitem{Kasuya:2006kj}
  S.~Kasuya and M.~Kawasaki,
  Phys.\ Rev.\  D {\bf 73}, 063007 (2006).

\bibitem{Chun:2006ss}
  E.~J.~Chun and H.~B.~Kim,
  JHEP {\bf 0610}, 082 (2006).

\bibitem{Pospelov:2007xh}
  M.~Pospelov and A.~Ritz,
  Phys.\ Lett.\  B {\bf 651}, 208 (2007).

\bibitem{Boyarsky1}  
  A.~Boyarsky, D.~Iakubovskyi, O.~Ruchayskiy and V.~Savchenko,
  Mon.\ Not.\ Roy.\ Astron.\ Soc.\  {\bf 387}, 1361 (2008).

\bibitem{Boyarsky2}
  A.~Boyarsky, D.~Malyshev, A.~Neronov and O.~Ruchayskiy, 
  Mon.\ Not.\ Roy.\ Astron.\ Soc.\  {\bf 387}, 1345 (2008).

\bibitem{Dodelson:1993je}
  S.~Dodelson and L.~M.~Widrow, Phys.\ Rev.\ Lett.\  {\bf 72}, 17 (1994).

\bibitem{Dolgov:2000ew}
  A.~D.~Dolgov and S.~H.~Hansen,
  Astropart.\ Phys.\  {\bf 16}, 339 (2002).

\bibitem{Asaka}
  T.~Asaka, M.~Laine and M.~Shaposhnikov,
  JHEP {\bf 0606}, 053 (2006) and JHEP {\bf 0701}, 091 (2007).

\bibitem{Yuksel:2007xh}
  H.~Yuksel, J.~F.~Beacom and C.~R.~Watson,
  Phys.\ Rev.\ Lett.\ {\bf 101} 121301 (2008).   

\bibitem{Palazzo:2007gz}
  A.~Palazzo, D.~Cumberbatch, A.~Slosar and J.~Silk,
  Phys.\ Rev.\  D {\bf 76}, 103511 (2007).

\bibitem{Shaposhnikov}
  M.~Shaposhnikov and I.~Tkachev, Phys.\ Lett.\  B {\bf 639}, 414 (2006).

\bibitem{Kusenko}
  A.~Kusenko, Phys.\ Rev.\ Lett.\  {\bf 97}, 241301 (2006).

\bibitem{Petraki:2007gq}
  K.~Petraki and A.~Kusenko,
  Phys.\ Rev.\  D {\bf 77}, 065014 (2008).

\bibitem{LeeWeinberg}
B.~W.~Lee and S.~Weinberg,
  Phys.\ Rev.\ Lett.\  {\bf 39}, 165 (1977).

\bibitem{KolbTurner}
E.~W.~Kolb and M.~S.~Turner, 
 \textit{The Early Universe}, Addison-Wesley (1990).

\bibitem{Giudice:2000ex}
G.~F.~Giudice, E.~W.~Kolb and A.~Riotto,
  Phys.\ Rev.\ D {\bf 64}, 023508 (2001).

\bibitem{Khalil:2002mu}
 e.g.,
  S.~Khalil, C.~Munoz and E.~Torrente-Lujan,  New J.\ Phys.\  {\bf 4} (2002) 27;
  C.~Pallis, Astropart.\ Phys.\  {\bf 21} (2004) 689;
  G.~B.~Gelmini and P.~Gondolo, Phys.\ Rev.\  D {\bf 74}, 023510 (2006).

\bibitem{Gelmini}
G.~Gelmini, S.~Palomares-Ruiz and S.~Pascoli,
  Phys.\ Rev.\ Lett.\  {\bf 93}, 081302 (2004).

\bibitem{Yaguna:2007wi}
C.~E.~Yaguna,
  JHEP {\bf 0706}, 002 (2007).

\bibitem{Gelmini:2008fq}
  G.~Gelmini, E.~Osoba, S.~Palomares-Ruiz and S.~Pascoli,
  arXiv:0803.2735 [astro-ph].

\bibitem{LowerBoundOnTR}
M.~Kawasaki, K.~Kohri and N.~Sugiyama,
  Phys.\ Rev.\ Lett.\  {\bf 82}, 4168 (1999)
  and Phys.\ Rev.\  D {\bf 62}, 023506 (2000);
K.~Ichikawa, M.~Kawasaki and F.~Takahashi,
  Phys.\ Rev.\ D {\bf 72}, 043522 (2005).

\bibitem{Bolz}
M.~Bolz, W.~Buchmuller and M.~Plumacher,
  Phys.\ Lett.\  B {\bf 443}, 209 (1998);
M.~Bolz, A.~Brandenburg and W.~Buchmuller,
  Nucl.\ Phys.\  B {\bf 606}, 518 (2001)
  [Erratum-ibid.\  B {\bf 790}, 336 (2008)] ;
F.~D.~Steffen,  JCAP {\bf 0609}, 001 (2006).

\bibitem{Affleck:1984fy}
I.~Affleck and M.~Dine,
  Nucl.\ Phys.\  B {\bf 249}, 361 (1985).

\bibitem{Dolgov:2002vf}    
E.~D.~Stewart, M.~Kawasaki and T.~Yanagida,
  Phys.\ Rev.\  D {\bf 54}, 6032 (1996);
B.~A.~Campbell, M.~K.~Gaillard, H.~Murayama and K.~A.~Olive,
  Nucl.\ Phys.\  B {\bf 538}, 351 (1999);
A.~D.~Dolgov, K.~Kohri, O.~Seto and J.~Yokoyama,
  Phys.\ Rev.\  D {\bf 67}, 103515 (2003).


\end{thebibliography}
\end{document}